\documentclass[aps,prl,twocolumn]{revtex4}
\usepackage{graphicx}
\usepackage{color}
\definecolor{darkred}{rgb}{0.75,0.0,0.0}
\newcommand\beq{\begin{equation}}
\newcommand\eeq{\end{equation}}
\newcommand\bear{\begin{eqnarray}}
\newcommand\eear{\end{eqnarray}}
\begin{document}

\title {Intrinsic Half-Metallicity in Modified Graphene Nanoribbons}

\author{Sudipta Dutta, Arun K. Manna and Swapan K. Pati}

\address{Theoretical Sciences Unit and DST Unit on Nanoscience
\\Jawaharlal Nehru Center for Advanced Scientific Research 
\\Jakkur Campus, Bangalore 560 064, India.}

\date{\today}

\begin{abstract}
\parbox{6in}

{We perform first-principles calculations based on density 
functional theory to study quasi one-dimensional edge-passivated 
(with hydrogen) zigzag graphene nanoribbons (ZGNRs) of various 
widths with chemical dopants, boron and nitrogen, keeping the 
whole system isoelectronic. Gradual increase in doping concentration 
takes the system finally to zigzag boron nitride nanoribbons (ZBNNRs). 
Our study reveals that, for all doping concentrations the systems 
stabilize in anti-ferromagnetic ground states. Doping concentrations 
and dopant positions regulate the electronic structure of the 
nanoribbons, exhibiting both semiconducting and half-metallic 
behaviors as a response to the external electric field. Interestingly, 
our results show that ZBNNRs with terminating polyacene unit exhibit 
half-metallicity irrespective of the ribbon width as well as applied 
electric field, opening a huge possibility in spintronics device 
applications.}  
\end{abstract}

\maketitle

\narrowtext

Nanomaterials of carbon, like nanotubes, fullerenes, etc., have 
been of great interest in condensed-matter and material science 
because of their novel low-dimensional properties \cite{Dresselhaus,Lakshmi}. 
Over past few decades, cutting edge research has been carried out for 
advanced device integration, exploring the electronic and mechanical 
properties of these systems. The recent addition in this journey is 
graphene: a strictly two-dimensional flat monolayer of carbon atoms 
tightly packed into a honeycomb lattice \cite{Novo1,Katsnelson}.  
Since its innovation \cite{Novo2,Novo3,Meyer}, it has made possible 
the understanding of various properties in two dimensions, by simple 
experiments and has opened up huge possibilities for electronic device 
fabrications \cite{Stankovich,Watcharotone,Novo4}. A large number of 
theoretical and experimental groups all over the world have gathered 
on this two dimensional platform to search for the "plenty of room" 
at this reduced dimension \cite{Neto1}. 

Electronic properties of low dimensional materials are mainly governed 
by their size and geometry. Recent experimental sophistications permit the 
preparation of finite size quasi one dimensional graphene, named as 
graphene nanoribbons (GNRs) of varying widths, either by cutting 
mechanically exfoliated graphenes and patterning by lithographic 
techniques \cite{Tapaszto,Datta} or by tuning the epitaxial growth 
of graphenes \cite{Berger1,Berger2}. Different geometrical 
terminations of the graphene monolayer give rise to two different 
edge geometries of largely varying electronic properties, namely, 
zigzag and armchair graphene. Several theoretical models, e.g., 
tight-binding model within Schrodinger \cite{Fujita,Nakada,Wakabayashi}, 
Dirac formalism for mass less fermions \cite{Brey,Abanin,Neto2}, density 
functional theory (DFT) etc. have been applied to explore the 
electronic and band structure properties of GNRs. There exists a 
few many-body studies, exploring the electronic and magnetic 
properties of these systems \cite{Sudipta1,Yang}.

\begin{figure}
\centering
\includegraphics[scale=0.25, angle=0] {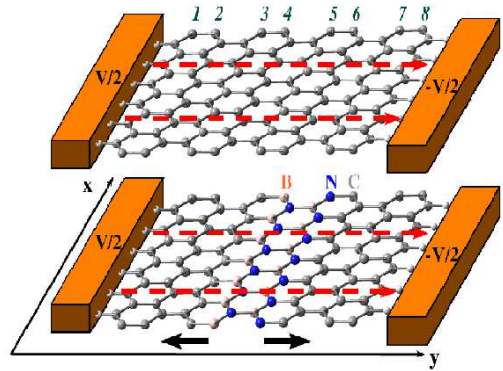}
\caption{(Color online) Schematic representation of the systems considered. 
The upper device is fabricated with hydrogen passivated  8-ZGNR (8 zigzag
chains) without any doping. The lower device depicts the doping positions
in 8-ZGNR, replacing two middle zigzag carbon chains by isoelectronic
boron-nitrogen chains. Doping concentration increases gradually along
the directions of solid arrows. Electric field (dashed arrows) has been
applied along the cross-ribbon width, which is the y-axis direction.}
\end{figure}

DFT studies suggest that, the anti-ferromagnetic quasi one-dimensional 
(1D) zigzag edge graphene nanoribbons (ZGNRs) show half-metallicity at 
a finite external electric field across the ribbon width within both 
local density approximation (LDA) \cite{Cohen} and generalized gradient 
approximation (GGA) \cite{Sudipta2}. Half-metallic materials show zero 
band gap for electrons with one spin orientation and insulating or 
semiconducting band gap for the other, resulting in completely spin 
polarized current. This interesting property has been observed in some 
materials like Heusler compounds \cite{Groot}, manganese perovskites 
\cite{Park}, metal-DNA complexes \cite{Sairam} etc. Modification of 
edges and systematic doping indeed change the electronic properties of 
GNRs and thus give new inroads in fine tuning of its band gap, controling 
the nature of the spin polarization  \cite{Hod,Kan,Gorjizadeh,Neto3}. 
Experimental progresses have enabled realization of such materials to 
a great extent \cite{Tapaszto,Datta,Mullen1,Mullen2,Delgado,Li}. 
In this letter, we focus on the effect of chemical doping on edge 
passivated ZGNRs. By chemical doping, we mean the replacement of carbon 
atoms by some other elements, keeping the system isoelectronic.

This can only be achieved if two carbon ($1s^{2} 2s^{2} 2p^{2}$) 
atoms in ZGNRs are replaced by one nitrogen ($1s^{2} 2s^{2} 2p^{3}$) and 
one boron ($1s^{2} 2s^{2} 2p^{1}$) atom. The effect of these chemical 
dopants has already been studied both experimentally and theoretically 
in case of nanotubes \cite{Rao}. It has been observed that, the ratio of 
carbon atoms to these dopants has significant impact on the magnetic 
behavior of the nanotubes and increse in boron and nitrogen concentration 
gradually opens up the gap between the valence and conduction band. Here 
we start with undoped hydrogen passivated 8-ZGNR (8 zigzag chains) and 
incorporate the dopants by replacing the two middle zigzag carbon chains 
by two zigzag boron-nitrogen (B-N) chains in such a way that, each C-C 
unit is replaced by one B-N unit, resulting in a situation where the 
middle of the ZGNR looks like as a polyborazine (Figure 1). We then 
systematically increase the doping concentration by replacing the two 
adjacent zigzag carbon chains, just next to the polyborazine at the middle 
and keep doing so upto the edges. Ultimately we end up at the zigzag boron 
nitrogen nanoribbon (ZBNNR), where all the C-C units are replaced by 
B-N pairs. Although the ZGNRs and ZBNNRs are isoelectronic, they exhibit 
completely different electronic properties. As like bulk hexagonal boron 
nitride, quasi one-dimensional ZBNNRs show wide gap insulating behavior 
\cite{Nakamura,Guo,Louie}. In this letter, we consider chemically doped
ZGNRs with varying widths and study their electronic properties with and
without external electric field.

\begin{figure}
\centering
\includegraphics[scale=0.28, angle=0] {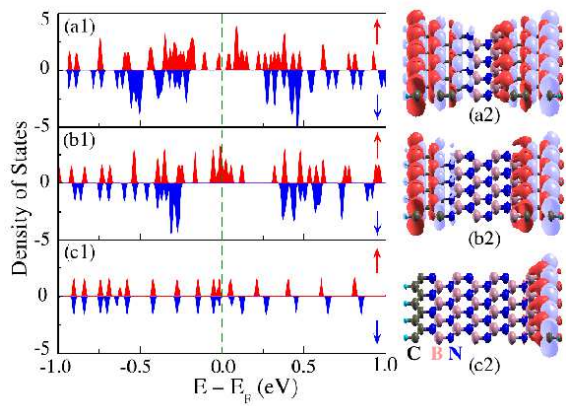}
\caption{(Color online) The Density of states as a function of energy, 
scaled with respect to the Fermi energy for the 8-ZGNR with (a1) 2, 
(b1) 4 and (c1) 6 zigzag carbon chains replaced by the zigzag 
boron-nitrogen chains at the middle of the nanoribbon. The spin 
density profile of these systems are shown in (a2), (b2) and (c2)
respectively.}
\end{figure}

We perform first principle periodic calculations based on density functional 
theory adopting the SIESTA package \cite{Soler}. Within generalized gradient 
approximation (GGA), considering Perdew-Burke-Ernzerhof (PBE) exchange and 
correlation functionals \cite{Burke} with a double zeta polarized (DZP) basis 
set, we have performed spin polarized calculations. The pseudopotentials 
are constructed from 3, 4 and 5 valence electrons for the B, C and N atoms, 
respectively. GGA approximation takes into account the semilocal exchange 
correlations which have significant impact on low-dimensional spin systems 
like GNRs and BNNRs. We have considered the nanoribbons translated infinitely 
along the x-axis, as shown in Figure 1. We consider the lattice vectors along 
the y and z-axis as 50.0 $\AA$ and 16.0 $\AA$  respectively, creating sufficient 
vacuum to avoid the interactions between the adjacent unit cells. We have 
considered a 400 Ry energy cutoff for a real space mesh size and a k-point 
sampling of 36 k-points, uniformly distributed along the 1 D Brillouin zone. 
We have considered nanoribbon widths of 8, 12 and 24 zigzag chains with 
varying concentration of doping and relax all of the structures with both 
ferromagnetic and anti-ferromagnetic spin orientations. A transverse electric 
field has been applied along the cross-ribbon direction, i.e., along the 
y-axis (see Figure 1) on the ground state structures to mimic the band gap 
engineering for both the spins.

All the systems show zero ground state magnetic moment. The band structure 
analysis shows that, the undoped ZGNRs are narrow gap semiconductors with 
similar gaps for up and down spins. On application of external electric 
field, one of the spin gaps closes at a certain critical field strength, 
making the system half-metallic. The direction of applied field governs 
the nature of the conducting spin channel \cite{Cohen,Sudipta2}. On the 
other hand, the ZBNNRs are known as wide gap insulators and the gap closes 
on application of electric field without any spin polarization of the 
current \cite{Guo}. These are two extreme cases. Gradual replacement of 
zigzag carbon chains at the middle of the ZGNRs with zigzag B-N chains 
takes the ZGNRs to ZBNNRs and the electronic and band structures change 
accordingly. In Figure 2, we have shown the density of states (DOS) of 
8-ZGNR with replacement of (a1) 2, (b1) 4 and (c1) 6 middle zigzag carbon 
chains. The undoped 8-ZGNR shows band gap of 0.5 eV for both the spins 
\cite{Sudipta2}. Replacement of 2 middle zigzag carbon chains gives 
non-zero DOS near the Fermi energy for one of the spins, reducing the gap 
significantly but keeping the other spin channel semiconducting. The 
system with 4 zigzag B-N chains at the middle makes one of the spin channels 
conducting. But there is an increase in the gap for the other spin channel.  
Interestingly, the system with 6 zigzag B-N chains at the middle and two 
zigzag carbon chains at the edges shows decrease in both the spin gaps, a 
distinct behavior different from the other two cases. The DOS analysis in 
first two cases suggest that the increase in doping concentration closes 
one of the spin gaps and opens up for the other spin channel. These 
observations clearly indicate a deviation from the symmetric spin 
distribution in bipartite ZGNRs.

\begin{figure}
\centering
\includegraphics[scale=0.30, angle=0] {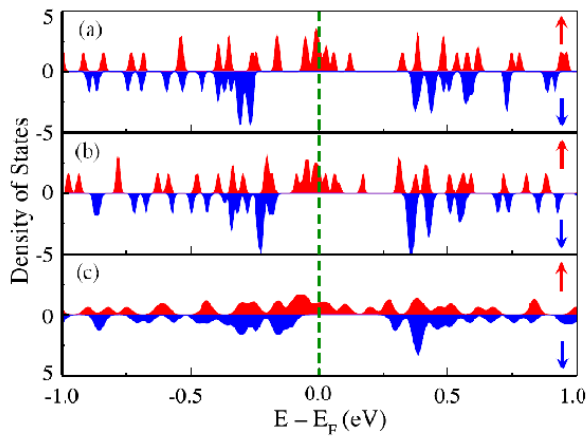}
\caption{(Color online) The density of states as a function of energy, 
scaled with respect to the Fermi energy for (a) 8, (b) 12 and (c) 24 
ZBNNRs with terminating polyacene unit.}
\end{figure}

The ground state of ZGNRs with zero magnetic moment, show antiparallel spin 
alignment between nearest neighbours, a typical nature of bipartite lattice 
which has earlier been noted \cite{Cohen,Sudipta2,Lieb}.  
Although all the atoms in one edge are aligned ferromagnetically, they show 
anti-ferromagnetic coupling within two edges. However, in the ZBNNRs, both the 
spins prefer to mix homogeneously throughout the whole lattice without any 
dominance of a particular spin on any site. Figure 2 shows the spin 
distributions of the intermediate systems, i.e., zigzag boron-nitrogen chain 
doped ZGNRs. The figure suggests that, whenever carbon atoms find nitrogen 
or boron as nearest neighbor, the bipartite symmetry breaks. Moreover, due 
to the Lewis acid character, the boron atoms pull the electrons from adjacent 
carbon atoms, resulting in a charge transfer from carbon to boron and creating 
a potential gradient across the ribbon width (see Figure 2). Boron and nitrogen 
face different spins on the adjacent carbon atoms in either side of the doped 
zigzag lines. Thus unsymmetrical spin diffusion on either side destroys the 
spin symmetry, which is reflected in the DOS calculations. The spin distribution 
of the system with 6 zigzag B-N chains at the middle and terminating 
polyacetylene chains (Figure 2 (c2)) shows localization of spins on one of the 
edges only, resulting in a completely inhomogeneous electronic structure 
resembling neither ZGNRs nor ZBNNRs. Note that, all the other doped systems 
are mixtures of polyborazine and polyacene units which are the constituents 
of ZGNRs and ZBNNRs respectively and thus exhibits mixed properties. Infact, 
carbon dopants in this system (ZBNNR with terminating polyacetylene chains) 
insert finite density of states for both the spins within the large 
insulating gap of ZBNNR. To estimate the contribution of the terminating 
carbon atoms, we perform calculations for projected density of states (pDOS), 
which indeed shows that, the DOS near Fermi energy come solely from the edge 
carbon atoms.  

\begin{figure}
\centering
\includegraphics[scale=0.5, angle=0] {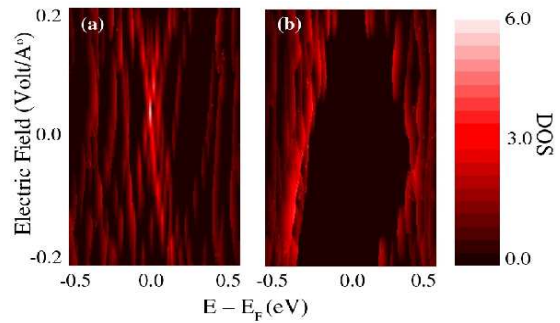}
\caption{(Color online) Contour plot of the density of states as function of
energy, scaled with respect to the Fermi energy for (a) conducting and (b)
insulating spin channels of 8-ZBNNR with terminating polyacene unit with
varying electric field.}
\end{figure}

pDOS analysis for all the systems show that the maximum contribution comes 
from the carbon atoms and the extent of their contributions reduce from the 
edges towards the middle. However, the carbons, attached to the boron atoms 
and subsequent carbon atoms on the same side of doped ZGNRs shows higher 
contributions than that of the other side, which is attached to the nitrogen 
atoms. This observation unambiguously proves that, the half-metallicity in 
boron-nitrogen doped ZGNRs originates from the charge transfer from the 
carbon to boron atoms. This can be further explained from the bare onsite 
Coulomb repulsion values of carbon ($\sim 9.66$ eV), boron ($\sim 8.02$ eV) and 
nitrogen ($\sim 14.46$ eV), which has been estimated as the difference between 
the ionization potential and electron affinity of the bare atoms. Due to 
reduced onsite Coulombic replulsion, the electrons prefer to flow from carbon 
to boron, whereas the charge transfer from carbon to nitrogen is hindered by 
the large Coulombic repulsion on nitrogen sites. Although these atomic values 
renormalize in case of complex systems like nanoribbons, their relative 
differences remain almost unaffected.

Calculations with varying ZGNR width with all possible doping concentrations 
result in similar observations. Our study reveals that the systems with two 
zigzag carbon chains on either edges and all zigzag B-N chains at the middle 
shows half-metallicity irrespective of the ribbon width, which has 
been shown in Figure 3. pDOS analysis for all the widths confirms the Lewis 
acid character of boron for the observed half-metallicity. The insulating 
gap for the other spin channel decreases with an increase in ribbon width, 
manifesting the characteristics of both ZGNRs and ZBNNRs in absence of doping 
\cite{Cohen,Sudipta2,Guo}. In Figure 4, we have shown the variation of the 
band gap for both the spin channels for 8-ZBNNR system with terminal polyacene 
units. Application of electric field on ZBNNRs with polyacene edges for all 
widths shows that, the half-metallic behavior sustains over even sufficiently 
large e-field strength. The magnitude of the semiconducting spin gap 
($> 0.3$ eV) suggests the possibilities of experimental realization of such 
half-metallic anti-ferromagnets even at room temperature ($0.025$ eV).

\begin{figure}
\centering
\includegraphics[scale=0.32, angle=0] {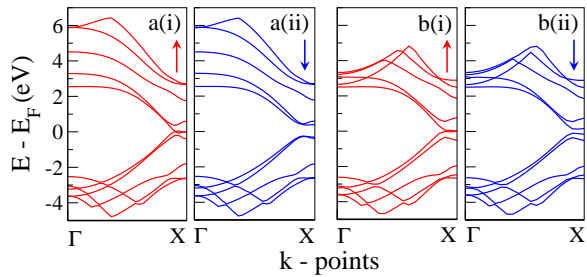}
\caption{(Color online) The spin polarized band structure (5 valence + 5
conduction) of 8-ZBNNR with terminating polyacene unit, as obtained from
(a) SIESTA and (b) VASP calculations for (i) up and (ii) down spins
separately. The Fermi energy  ($E_{F}$) is set to zero.}
\end{figure}

We have also verified that, the intrinsic half-metallicity as obtained for
ZBNNRs with terminating polyacene unit from SIESTA with localized basis remains 
qualitatively same with extended plane augmented wave \cite{PAW} basis 
within DFT package VASP \cite{vasp}. In Figure 5, the spin polarized band structure
results for 8-ZBNNR from both SIESTA and VASP at the same level of calculations 
are shown. As can be seen, although the gap for the insulating spin
channel decreases for extended basis states, it remains far above the room
temperature, highlighting our claim for intrinsic half-metallicity in ZBNNRs.

In conclusion, we have studied the ZGNRs with systematic chemical doping, 
replacing the zigzag carbon chains with zigzag boron-nitrogen chains in 
the middle of the nanoribbons, keeping the systems isoelectronic with 
the undoped ZGNRs to search for better half-metallic materials. Our study 
suggests that, the nanoribbons with terminating polyacene unit and all 
zigzag boron-nitrogen chains at the middle has the potential to act as a 
half-metallic anti-ferromagnet for all widths. We have identified the Lewis 
acid character of boron, that results in charge transfer form carbon to 
boron to invoke half-metallicity in graphene based materials. This property 
sustains over a large external electric field strength even at room 
temperature, opening huge possibilities for the realization of graphene 
based spintronic devices with complete control over the spins. 

S.D. and A.K.M. acknowledge CSIR, Govt. of India for research fellowship 
and S.K.P. acknowledges CSIR and DST, Govt. of India and AOARD, US Airforce 
for research funding.

\end{document}